# Orbital Selective Magnetism in the Spin-Ladder Iron Selenides Ba$_{1-x}$K$_x$Fe$_2$Se$_3$

J. M. Caron[1], J. R. Neilson[1,‡], D. C. Miller[1], K. Arpino[1], A. Llobet[2], and T. M. McQueen[1,†]

[1] Institute for Quantum Matter, Department of Chemistry, and Department of Physics and Astronomy, The Johns Hopkins University, Baltimore, MD 21030
[2] Los Alamos National Laboratory, Lujan Neutron Scattering Center, MS H805, Los Alamos, NM 87545

Here we show that the 2.80(8) $\mu_B$ Fe$^{-1}$ block antiferromagnetic order of BaFe$_2$Se$_3$ transforms into stripe antiferromagnetic order in KFe$_2$Se$_3$ with a decrease in moment to 2.1(1) $\mu_B$ Fe$^{-1}$. This reduction is larger than expected from the change in electron count from Ba$^{2+}$ to K$^+$, and occurs with the loss of the displacements of Fe atoms from ideal positions in the ladders, as found by neutron pair distribution function analysis. Intermediate compositions remain insulating, and magnetic susceptibility measurements show a suppression of magnetic order and probable formation of a spin-glass. Together, these results imply an orbital-dependent selection of magnetic versus bonded behavior, driven by relative bandwidths and fillings.

The origin of high-temperature superconductivity remains controversial.[1] In the case of iron-based superconductors, even the origin of magnetic and metallic ground states of non-superconducting parent compounds is contested. Theories range from the multiband character and nesting of the Fermi surface[2,3] and that iron pnictides are Hund's metals,[4] to proposals that the compounds are more directly related to the cuprates.[5,6] Further complication arises from the varied behavior of different iron compounds, including whether magnetic order competes[7,8] or coexists[9-11] with superconductivity, as well as the presence or absence of structural distortions,[12] and their relation to magnetic order.[13] Orbital-selection is critical to understanding the nature of these magnetic and structural states in iron-based superconductors, independent of the choice of theory.[14-16]

The spin-ladder compounds $A$Fe$_2X_3$ ($A$ = K, Rb, Cs, or Ba and $X$ = Chalcogenide) are structurally related to the iron superconductors,[17-19] and are built of double-chains ('ladders') of edge-sharing [Fe$X_4$] tetrahedra. The $A$ cations separate the ladders to form the three-dimensional structure [Fig. 1(a)]. Compared to the layered iron-based superconductors, the $A$Fe$_2X_3$ compounds have significantly decreased bandwidths from the reduced dimensionality [Fig. 1(b)]. Consequently, these compounds provide a unique opportunity to reveal the interplay between structure and magnetism in [Fe$X_4$]-based materials without the complication of metallic behavior. Here we report a systematic study of charge-doping of the spin-ladder compound Ba$_{1-x}$K$_x$Fe$_2$Se$_3$ using resistivity, magnetic susceptibility, synchrotron powder x-ray diffraction (PXRD) and neutron powder diffraction (NPD) with pair-distribution function analysis (PDF). The data intimate a relationship between local Fe–Fe displacements and magnetic order, driven by a shift in electronic state from unpaired and magnetic to paired and bonded as a function of electron count. Further, these results imply an orbitally-dependent preference for magnetism, suggesting that such relationships are pertinent to iron-based superconductivity.

Samples were made as previously reported.[20,21] All samples were prepared and handled inside an argon-filled glovebox. NPD data were collected at temperatures between 5 K and 300 K on polycrystalline KFe$_2$Se$_3$ loaded in a vanadium can using the time-of flight High Intensity Powder Diffractometer (HIPD) at the Lujan Center, Los Alamos Neutron Science Center, Los Alamos National Laboratory. During shipment and loading, a portion decomposed into FeSe and KSe$_2$ (volatile). Thus, a FeSe impurity phase (< 10 wt%), not seen in PXRD data of the same sample batch, was included in the neutron refinements. PDF analysis was performed on the total neutron scattering data and $G(r)$ were extracted with $Q_{max}$ = 29 Å$^{-1}$ using PDFgetN.[22] Refinements to the PDF were performed using PDFgui[23] after defining instrumental resolution parameters from data collected on polycrystalline Si ($Q_{broad}$ = 0.041 $Q_{damp}$ = 0.015). PXRD data were collected at 300 K on beamline 11-BM at the Advanced Photon Source. Rietveld analyses were performed using FullProf,[24] and GSAS/EXPGUI[25]. The magnetic structure was solved by representational analysis as implemented by SARA$h$ and BASIREPS/FullProf.[24]

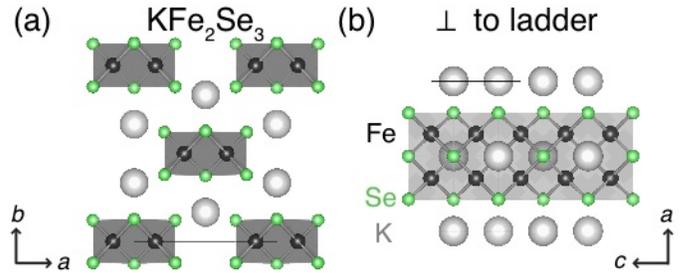

Figure 1 (Color Online). Schematic of KFe$_2$Se$_3$ (space group $Cmcm$) showing (a) stacking of [Fe$_2$Se$_3$] ladders and (b) the perspective perpendicular to the plane of the ladders.

Electrical resistivity and dc magnetization measurements were performed between 1.8 K and 300 K using a Quantum Design Physical Properties Measurement System. For resistivity, platinum wires were attached to sintered polycrystalline pellets using silver paste and dried in a desiccator. Magnetization measurements were carried out at $\mu_0H$ = 1 T and 2 T, and with $\chi \approx \Delta M/\Delta H = [M_{2T}-M_{1T}]/[1T]$.

Resistivity data, normalized to $\rho_{300K}$, are shown in Fig. 2(a). While substitution of Ba by K formally oxidizes iron from 2+ to 2.5+, all samples are insulating, with the resistivities increasing exponentially with decreasing temperature. The temperature dependence exhibits one-dimensional variable range hopping (VRH) behavior, given by $\ln(\rho/\rho_{300K}) \propto (A^2 T_0'/T)^{1/2}$, where $T$ is the temperature, $A$ is a constant, and $T_0'$ is the density of states in the localization length.[26] For most samples, $\log(\rho/\rho_{300\ K})$ vs. $T^{-1/2}$ is linear over the entire measurement range, but for $x$ = 0.3, 0.9, and 1.0, there are two discrete regions of linearity, with crossovers at $T_{cross}$ = 85(5), 70(5), and 170(5) K, respectively.

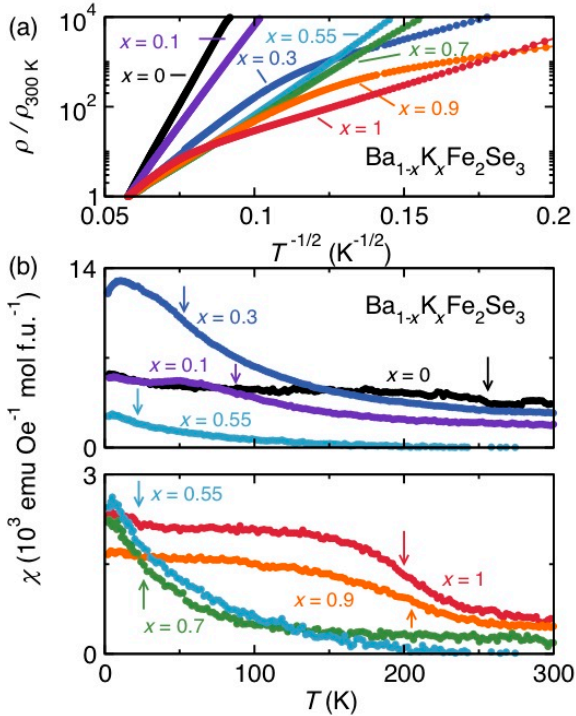

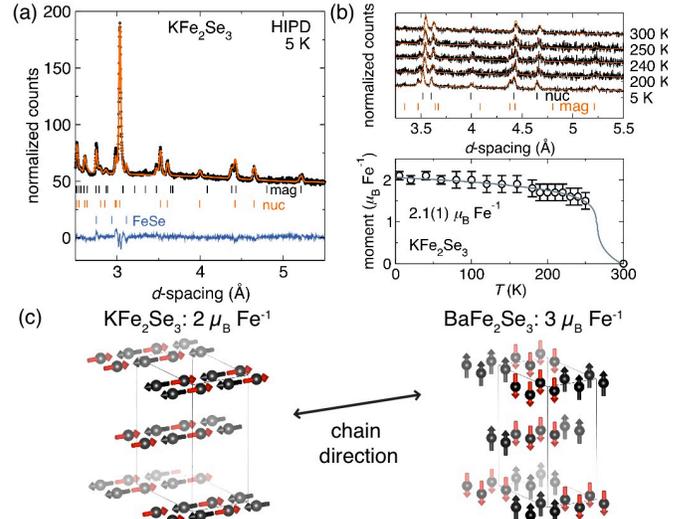

Figure 2 (Color Online). (a) Normalized electrical resistivity on polycrystalline pellets of $Ba_{1-x}K_xFe_2Se_3$ highlighting one-dimensional variable range hopping behavior. (b) The temperature dependences of the magnetic susceptibilities illustrate the initial suppression of magnetic order (indicated by arrows) with increasing $x$, followed by a reappearance of antiferromagnetic order.

To understand if the crossover behavior is related to magnetic order, as in orthorhombic iron-arsenides,[27,28] the magnetic susceptibilities of $Ba_{1-x}K_xFe_2Se_3$ are shown in Fig. 2(b). For $x$ = 0, we previously showed that there is a change in slope at $T_N$ = 256 K, corresponding to the formation of magnetic order.[20] On increasing $x$, the temperature of the slope change quickly, and drops and broadens. This observation is not surprising given that iron-deficient $BaFe_{1.79(2)}Se_3$ samples (nominally $Fe^{2.23+}$) have spin-glass behavior at $T$ < 50 K,[29] and imply that at intermediate $x$, $Ba_{1-x}K_xFe_2Se_3$ also exhibits spin-glass behavior. Long-range antiferromagnetic order reappears for $x \geq 0.9$; the apparent diffuse nature of the transition in magnetization is attributed to crystalline anisotropy.[30]

From these data, there is no obvious relation between the resistivity crossovers and magnetic susceptibility: for $x$ = 0.3, $T_{cross}$ = 85(5) K, but spin-glass formation does not occur until $T$ < 50 K, whereas for $x$ = 0.9, $T_{cross}$ = 70(5) K, but magnetic order sets in at a significantly higher temperature, $T_N \approx 250$ K. This suggests that the resistivity crossovers arise from some other effect. One possibility is changes in iron deviation deficiency, but our PXRD Rietveld refinements show no from the nominal stoichiometries from loss of K or Fe. More likely is that all samples do have the resistivity crossovers, but that for some the crossovers occur outside our measurement range. Further studies on single-crystal specimens are needed to identify whether the crossover behavior is present in all specimens, and whether they are magnetic in origin.

Figure 3 (Color Online). (a) NPD data (black circles) of $KFe_2Se_3$ and Rietveld refinement of the best magnetic structure. (b) Temperature dependent NPD data illustrate the growth of the magnetic Bragg peaks on cooling. The magnetic moment saturates at 2.1(1) $\mu_B$ $Fe^{-1}$. (c) Schematic of the stripe magnetic structure of $KFe_2Se_3$ compared to the block magnetic structure of $BaFe_2Se_3$.[20]

Table I: Crystal structure parameters of $KFe_2Se_3$ at 5 K. Space group: $Cmcm$, $a$ = 9.30(7) Å $b$ = 11.41(8) Å $c$ = 5.60(4) Å. The magnetic structure is described by a propagation vector of $\vec{k} = \langle \frac{1}{2}, \frac{1}{2}, 0 \rangle$ with $\Gamma_{irrep} = \Psi_3 - \Psi_6 + \Psi_9 - \Psi_{12}$ and a Fourier coefficient giving a moment of 2.1(1) $\mu_B$ $Fe^{-1}$.

| Site | x | y | z | $U_{iso}$ (Å$^2$) |
|---|---|---|---|---|
| K | 0.5 | 0.1667 | 0.25 | 0.013 |
| Fe | 0.352(2) | 0.5 | 0 | 0.003 |
| Se1 | 0 | 0.127(2) | 0.25 | 0.004 |
| Se2 | 0.212(2) | 0.379(2) | 0.25 | 0.004 |

To investigate the apparent suppression and reappearance of long-range magnetic order, the magnetic structure for $x$ = 1 ($KFe_2Se_3$) was solved using NPD. The best fit to data at $T$ = 5 K is shown in Fig. 3(a). Magnetic Bragg peaks appear on cooling [Fig. 3(b)] and are indexed as satellites to the $Cmcm$ unit cell[17] with a magnetic propagation vector of $\vec{k} = \langle \frac{1}{2}, \frac{1}{2}, 0 \rangle$. Representational analysis describes four irreducible representations, each spanned by 12 basis vectors corresponding to each unique Fe site. Trial magnetic structures were tested by Rietveld refinement to the data at 5 K, assuming all basis vectors point along the same crystallographic axis and are multiplied by a Fourier coefficient with the same magnitude. Only one configuration (Table I) describes the observed reflections ($R_{mag}$ = 15.4% vs. 28.9% for next best, 90° bank). The Fourier coefficient was refined as a function of temperature [Fig. 3(b)] and reveals a transition to a long-range magnetic state at $T_N \approx 250$ K, with the moment saturating at 2.1(1) $\mu_B$ $Fe^{-1}$. The magnetic structure of KFe$_2$Se$_3$ [Fig. 3(c)] is stripe antiferromagnetic order, with spins oriented in-chain, parallel to the chain direction. The stripe order is analogous to the magnetic structure observed in many parent pnictides such as BaFe$_2$As$_2$ and LaFeAsO.[31,32] This contrasts with the block antiferromagnetic order that we found in BaFe$_2$Se$_3$ with spins oriented perpendicular to the chain direction.[20] The block order has no direct analogs among known iron superconductors, but is closely related to the magnetic order of $A_x$Fe$_{2-y}$Se$_2$.[9]

In BaFe$_2$Se$_3$, local Fe–Fe displacements and magnetic order are coupled.[20,33] To determine if similar displacements are present in KFe$_2$Se$_3$, PDF analysis of KFe$_2$Se$_3$ was carried out at $T = 5$, 100, and 300 K. At all temperatures, the PDFs are well described by the average $Cmcm$ crystallographic structure [Fig. 4(a)], with a single Fe–Fe distance (2.83 Å) along the chain direction. The poorest fit is to the $T = 300$ K PDF, and puts an upper bound on local Fe–Fe distances from 2.75 Å to 2.87 Å ($\Delta \leq 0.12$ Å). This is substantially less than in BaFe$_2$Se$_3$,[20] where Fe–Fe distances of 2.62 Å and 2.83 Å ($\Delta = 0.21$ Å) are found [Fig. 4(b)]. These data show that the large local Fe–Fe displacements seen in BaFe$_2$Se$_3$ are much reduced or absent in KFe$_2$Se$_3$.

The properties of Ba$_{1-x}$K$_x$Fe$_2$Se$_3$ are summarized in Fig. 5. For $x > 0$, block antiferromagnetic order (AFM-B) is suppressed, indicated by the rapid decrease in $T_{mag}$. This is followed by the appearance of stripe antiferromagnetic order (AFM-S) for $x > 0.9$. At intermediate $x$, long-range magnetic order is likely replaced by a spin-glass state. Structural parameters from Rietveld refinement of PXRD data using the $Pnma$ space group generates long and short Fe–Fe distances [Fig. 5b, closed diamonds and triangles, respectively]. The difference between the long and short Fe–Fe separations remains nearly constant with $0 < x \leq 0.5$, while the average Fe–Fe distance [Fig. 5(b), open red circles] steadily increases. For $x > 0.5$, the magnitude of the average Fe–Fe displacements are reduced. Local Fe–Fe displacements, determined by PDF analysis [closed squares, Fig. 5(b)], confirm distinct short and long distances for $x = 0$ that are dramatically reduced or vanish when $x = 1$. Further support for the loss of Fe–Fe displacements comes from the change in space group from $Pnma$ to $Cmcm$ around $x = 0.9$. Fe–Fe displacements are allowed in $Pnma$, but not in $Cmcm$. Together, these data imply a large change in local Fe–Fe distances and exchange interactions across the series.

These structural changes also track with changes in one-dimensional VRH seen in resistivity. For $x < 0.7$, we observe a linear increase in $[(k_B T_0)^{-1}]$ (with $T_0$ defined as $A^2 T_0'$ and fit to the high-$T$ regions). The value of $[(k_B T_0)^{-1}]$ is proportional to the product of the localized density of states and the localization length.[26] This increase is consistent with the addition of carriers on doping, assuming a constant localization length. For $x \geq 0.9$, the trend in $[(k_B T_0)^{-1}]$ with $x$ inverts, indicating either a reduction in the density of states, or increased localization, of carriers. Such an inflection implies a non-trivial rearrangement of the electronic structure.

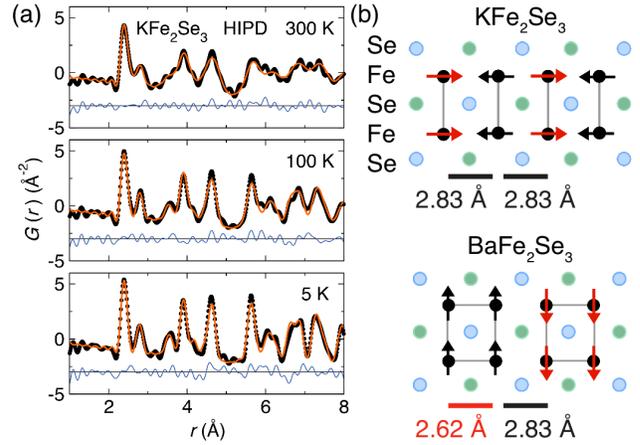

Figure 4 (Color Online). (a) PDFs of KFe$_2$Se$_3$ at $T = 5$, 100, and 300 K (black circles) with fits to the crystallographic $Cmcm$ structure. (b) Representations of a single ladder illustrate the changes in local structure and magnetism with substitution of the $A$ site in $A$Fe$_2$Se$_3$.

Explanation of these data requires consideration of the multiband electronic structure, derived from the five $3d$ orbitals near the Fermi level. The compounds studied here are one-dimensional and all are insulating. Therefore, all bands that would cross $E_F$ must be gapped by some mechanism. The local bonding geometry and electron count are comparable to the two-dimensional iron superconductors. As such, the bands that must be gapped are similar; in this case, four partly-filled metal $d$-orbital-derived bands.[29,34]

Gap formation can arise from on-site repulsion (Hubbard $U$) or by the formation of non-magnetic order, such as a charge density wave (CDW).[35-37] Localization by a Hubbard $U$ to form magnetic states is favored for narrow bands near half-filling. For the empirically observed 2.80(8) $\mu_B$ Fe$^{-1}$ observed in BaFe$_2$Se$_3$, this implies that $U$ gaps three bands, while a density wave or some more complex order opens a gap in the fourth. Such a density wave, if it produces a CDW, would then explain the two distinct Fe–Fe distances along the chain in BaFe$_2$Se$_3$. This distortion in turn causes variations in NN exchange, resulting in the block magnetic structure.

In KFe$_2$Se$_3$, there are $0.5 e^-$ per Fe fewer electrons, which, in the simplest case, results in quarter-filling of one of the three bands originally split by $U$ [Fig. 5(c)]. If $U$ was still responsible for the splitting of this third band, then a magnetic moment of 2.5 $\mu_B$ Fe$^{-1}$ would be observed. However, NPD shows only 2.1(1) $\mu_B$ Fe$^{-1}$. This implies that one of the three magnetic bands in BaFe$_2$Se$_3$ is instead gapped in a non-magnetic fashion in KFe$_2$Se$_3$ (the effect of $U$ is reduced away from half-filling). The switch to non-magnetic behavior then either produces a second CDW' with periodicity/phase different than the first, or some more complex density wave (valley density wave,[38] bond-order wave,[39] etc.). In either case, there is no longer a single CDW with a periodicity of $2d_{Fe-Fe}$, resulting in a reduction or elimination of the Fe–Fe displacements, as observed for KFe$_2$Se$_3$.

Such a change also explains the suppression and reappearance of magnetic order due to variation in local exchange from Fe–Fe displacements. Such disorder reduces spin-freezing temperatures, even when the individual exchange interactions remain strong.[40] The crystal structures of $Ba_{1-x}K_xFe_2Se_3$ also support this model; depopulation of the magnetic band for $x \leq 0.5$ leaves the $2d_{Fe-Fe}$ CDW intact, retaining the long and short Fe–Fe distances [Fig. 5(b)]. For $x > 0.5$, the shrinking magnitude of the Fe–Fe displacements indicates the switch to more complex non-magnetic density wave behavior.

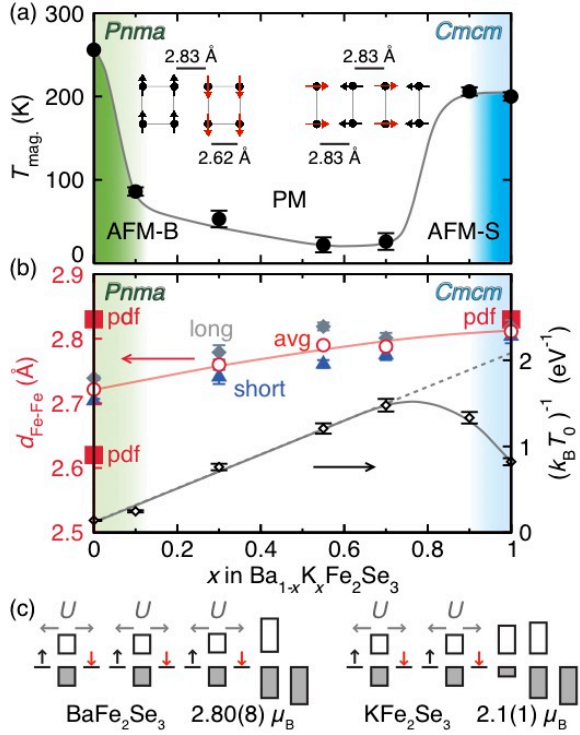

Figure 5 (Color Online). Phase diagram of $Ba_{1-x}K_xFe_2Se_3$. (a) As $x$ increases, block antiferromagnetic order (AFM-B) is suppressed, but for $x \geq 0.9$, stripe antiferromagnetic order (AFM-S) appears. (b) The average Fe–Fe distance at 300 K steadily increases (open circles), while the Fe–Fe displacements, determined by Rietveld (long, closed diamonds; short, closed triangles) and PDF analysis of $KFe_2Se_3$ (closed squares), vanish. Concomitantly, a linear increase in $[(k_BT_0)^{-1}]$ is consistent with the addition of hopping carriers. Above $x > 0.7$, a downturn indicates either a reduction in the number of, or an increase in localization of, carriers. (c) Orbital selective magnetism in $Ba_{1-x}K_xFe_2Se_3$: $BaFe_2Se_3$ has four half-filled $d$ bands, three narrow and one dispersed. While on-site repulsion ($U$) gaps the narrow bands, the dispersed band is gapped by a density wave. In $KFe_2Se_3$, depopulation reduces the effect of $U$ on one of the narrow bands resulting in its switch to non-magnetic behavior.

This switch from three to two bands gapped by $U$ also means that the orbitals contributing to the magnetic moment changes (as each band has a different set of orbital contributions), implying an orbital selection of magnetism. This change explains the switch in moment direction, since the shape of the electron density no longer contains contributions from the third band. Further, this model explains the inflection of $[(k_BT_0)^{-1}]$: incremental increases in charge-doping beyond $x > 0.7$ do not provide additional carriers; instead, the added carriers are localized by a density wave.

In conclusion, the competition of bandwidth and Hubbard $U$ for each $d$ band in $Ba_{1-x}K_xFe_2Se_3$ brings an orbital selectivity to the magnetic state. The change to antiferromagnetic stripe order is a natural consequence of the more uniform nearest-neighbor exchange interactions, arising from a change in underlying density wave(s). The insulating behavior results from different bands being gapped by different mechanisms; some are magnetic, some are not. Our results imply that small changes in band filling have a large effect on the nature of each $d$ band in iron-based materials. Further work is needed to identify the specific change in orbital contributions in $Ba_{1-x}K_xFe_2Se_3$. We speculate that this sensitivity, combined with interactions between bands, can give rise to exotic spin-singlet states and may explain the high-temperature superconductivity in these systems.

This research is supported by the U.S. Department of Energy, Office of Basic Energy Sciences, Division of Materials Sciences and Engineering under Award DE-FG02-08ER46544. This work has benefited from the use of HIPD at the Lujan Center at Los Alamos Neutron Science Center, funded by DOE Office of Basic Energy Sciences. Los Alamos National Laboratory is operated by Los Alamos National Security LLC under DOE Contract DE-AC52-06NA25396. This research has also benefited from the use of the Advanced Photon Source at Argonne National Laboratory supported by the U. S. Department of Energy, Office of Science, Office of Basic Energy Sciences, under Contract No. DE-AC02-06CH11357.

[†] E-mail: mcqueen@jhu.edu

[‡] E-mail: jneilso2@jhu.edu

for 14 hrs. The pre-reacted powders were then ground, pressed into pellets, and reheated in alumina crucibles in evacuated quartz tubes at 765°C (Ba) and 700°C (K) for 12 hr at a time, followed by quenching in water (K only). Re-grindings and reheatings were performed until each sample was phase pure as determined by laboratory x-ray diffraction (6-10 repetitions). Intermediate compositions $Ba_{1-x}K_xFe_2Se_3$, where $x$ = 0.1, 0.3, 0.55, 0.7, and 0.9, were prepared by mixing the appropriate ratio of the end members and heating once, as pellets, in alumina crucibles in evacuated quartz tubes at 700 °C followed by quenching in water.